\documentclass[review,3p]{elsarticle}

\usepackage{verbatim}
\usepackage{lineno,hyperref,graphicx}
\usepackage{morefloats}
\usepackage[usenames, dvipsnames]{color}
\usepackage{soul}
\usepackage{changes}
\usepackage{perpage}
\usepackage{multirow}
\usepackage[hang,tight]{subfigure}
\MakePerPage{footnote}
\modulolinenumbers[1]

\journal{Nuclear Instruments and Methods in Physics Research Section A}

\begin{document}

\begin{frontmatter}

\title{A GEM readout with radial zigzag strips and linear charge-sharing response}
\author[FIT]{Aiwu Zhang\corref{cor1}}
\cortext[cor1]{Corresponding author: Tel. +1(631)344-5483; Email \href{mailto:azhang@bnl.gov}{\textbf{azhang@bnl.gov}}. Now with Brookhaven National Laboratory.}
\author[FIT]{Marcus Hohlmann\corref{}}
\author[BNL]{Babak Azmoun}
\author[BNL]{Martin L. Purschke}
\author[BNL]{Craig~Woody}

\address[FIT]{Department of Physics and Space Sciences, Florida Institute of Technology, Melbourne, FL 32901, USA}
\address[BNL]{Physics Department, Brookhaven National Laboratory, Upton, NY 11973, USA}

\begin{abstract}
We study the position sensitivity of radial zigzag strips intended to read out large GEM detectors for tracking at future experiments. Zigzag strips can cover a readout area with fewer strips than regular straight strips while maintaining good spatial resolution. Consequently, they can reduce the number of required electronic channels and related cost for large-area GEM detector systems. A non-linear relation between incident particle position and hit position measured from charge sharing among zigzag strips was observed in a previous study. We significantly reduce this non-linearity by improving the interleaving of adjacent physical zigzag strips. Zigzag readout structures are implemented on PCBs and on a flexible foil and are tested using a 10 cm $\times$ 10 cm triple-GEM detector scanned with a strongly collimated X-ray gun on a 2D motorized stage. Angular resolutions of \mbox{60-84 $\mu$rad} are achieved with a \mbox{1.37 mrad} angular strip pitch at a radius of 784~mm. On a linear scale this corresponds to resolutions below \mbox{100 $\mu$m}. 
\end{abstract}

\begin{keyword}
GEM; Zigzag readout strip; Flex readout foil; Linear response; Spatial resolution
\end{keyword}

\end{frontmatter}

\section{Introduction} \label{Introduction}
The concept of zigzag-shaped readout pads was first proposed for gaseous time projection chambers (TPC) in the 1980s in order to reduce the number of electronic channels required to read out the detector~\cite{1985ZZ}. Later, MWPCs and GEMs were read out with parallel zigzag strips and good spatial resolutions were achieved~\cite{1997ZZ, 2005ZZ}. This was confirmed in more recent studies which showed that the spatial resolution of a small GEM detector with parallel zigzag strip or zigzag pad readout can approach 70 $\mu$m~\cite{2012ZZ, 2016ZZ}. We subsequently introduced radial zigzag strips to read out trapezoidal large-area GEM detectors~\cite{FITZZ2}, which are intended for tracking systems at future experiments, e.g.\ forward tracking at the electron ion collider (EIC)~\cite{EICWhite}. Radial zigzag strips can precisely measure the $\phi$ coordinates of incident particles in order to track them and to determine their transverse momenta in a solenoidal field.

In our previous studies~\cite{2016ZZ, FITZZ2}, we observed a non-linear relation between incident particle position and hit position measured from charge sharing among radial zigzag readout strips. This paper aims at quantifying the non-linear response of our previous zigzag designs and to demonstrate an improved zigzag design that has a linear response. A linear response ensures the accuracy of hit position measurements without the need for any corrections. For this purpose, six readout boards with different geometrical zigzag strip structures are produced and tested using a 10 cm $\times$ 10 cm triple-GEM detector on a 2D motorized stage. The incident particle position is defined by a highly collimated X-ray beam (140 $\mu$m $\times$ 8 mm collimator slit) in these measurements.

\section{Zigzag readout strip designs and test boards} \label{theZigzag}
\begin{figure}[h]
  \centering
  \includegraphics[width=8cm]{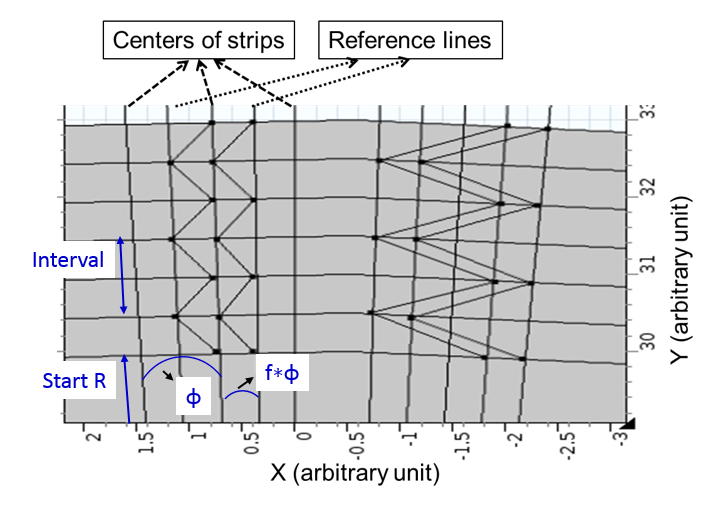}
  \vspace{-0.5cm}
  \caption{A sketch of radial zigzag strip designs indicating the four parameters that define the strip geometry. Here $\phi$ is the angular strip pitch and $f$ is the fractional strip pitch parameter that determines the strip shape. The tips of the left strip do not exceed the two reference lines, while the tips of the right strip reach the centers of its neighboring strips.}
  \label{zigzagSketch}
\end{figure}

As shown in Fig.~\ref{zigzagSketch}, zigzag strips can be designed by connecting certain points on strip center lines and reference lines in certain patterns. Four parameters are used to calculate the coordinates of these points: the starting radius of the strips, the period of the zigzag structure in the R direction (a fixed number of 0.5~mm in our studies), the $\phi-$angle pitch between strips (1.37 or 4.14 mrad in our studies), and a fraction $f$ of the angle pitch which defines the reference lines and determines the width, space, and interleaving of the zigzag strips~\cite{2015IEEE}. There are two ways to use these points. In the first design method shown on the left in Fig.~\ref{zigzagSketch}, a strip is outlined by points on a strip center line and two reference lines near it. In the second design method shown on the right in Fig.~\ref{zigzagSketch}, a strip is outlined by points on the two center lines of neighboring strips and on the two reference lines. We define the ``interleaving'' quantitatively as the distance of overlap between two adjacent strips in the azimuthal direction divided by the strip pitch. Then this second design has a ``100\% interleaving'' since the tips on one zigzag strip reach the centers of its two neighboring strips; this is expected to give better charge sharing and a more linear position response.

\begin{figure}[h]
  \begin{center}
     \subfigure{
          \label{zigzag1:left}
          \begin{minipage}[b]{0.5\textwidth}
              \centering
              \includegraphics[width=6cm,height=6cm]{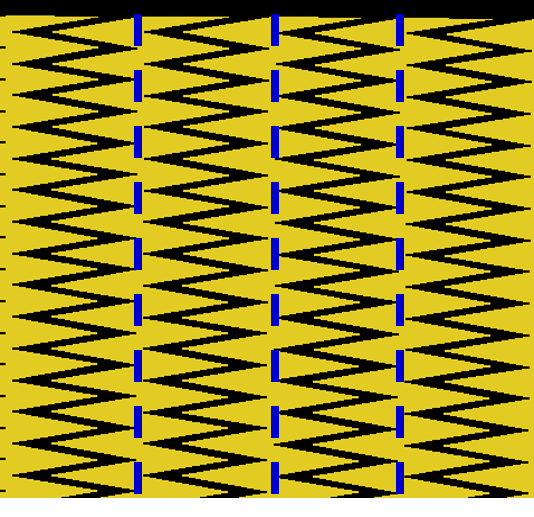}
          \end{minipage}}
     \subfigure{
          \label{zigzag1:right}
          \begin{minipage}[b]{0.5\textwidth}
              \centering
              \includegraphics[width=6cm,height=6cm]{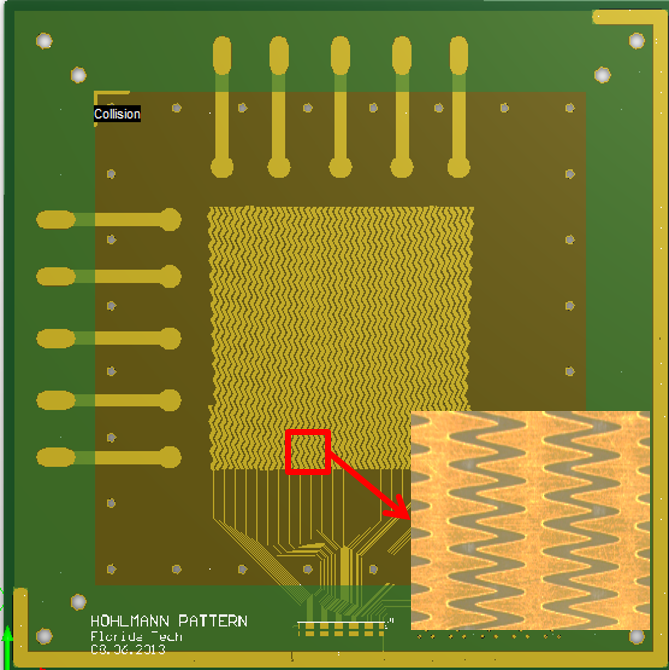}
          \end{minipage}}\newline%
    \vspace{-1cm}
    \caption{Left: The original zigzag strip design that was used in our previous study~\cite{FITZZ2}. The dashed blue lines represent the center lines of each zigzag strip. The angle pitch is 1.37 mrad, the period of the zigzag structure in the R direction is 0.5 mm and the fraction parameter $f$ is 0.75. Right: The full readout PCB design created in Altium Designer for 48 zigzag strips with this design. The strips have radii from 1420 to 1520 mm. The inset picture shows a region from the actually produced PCB, which demonstrates that the physical strips have a ``spine'' along their centers.}
    \label{zigzag1}
  \end{center}
\end{figure}

Fig.~\ref{zigzag1}~(left) shows the zigzag design produced with the first design method that was used in our previous study~\cite{FITZZ2}. The angle pitch was 1.37 mrad, the period of the zigzag structure in the R direction was 0.5~mm and the fractional strip pitch parameter was $f = 0.75$. Two printed circuit boards (PCBs) with this design were produced by industry (Fig.~\ref{zigzag1}~right): one with 48 strips of radii from 1420 to 1520 mm (``ZZ48 board''), the other with 30 strips of radii from 2240 to 2340 mm (``ZZ30 board''). The strip length on both boards was 10 cm except for strips near the edges of the active area and the strips on each board covered an area of approximately 10 cm $\times$ 10 cm. As can be seen in the inset in Fig.~\ref{zigzag1}~(right), the physical strips on the manufactured boards had a ``spine'' along the center of each strip due to low manufacturing precision in the etching of sharp points and corners. Also the space between strips turned out to be wider than what had been designed. No effort was made to further improve the quality of these two boards.

\begin{figure}[h]
\vspace{-1cm}
  \begin{center}
     \subfigure{
          \label{zigzag2:left}
          \begin{minipage}[b]{0.5\textwidth}
              \centering
              \includegraphics[width=6cm,height=6cm]{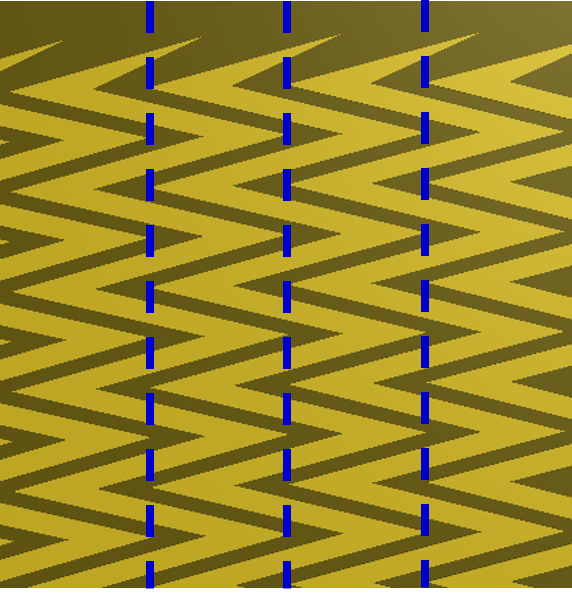}
          \end{minipage}}
    \subfigure{
          \label{zigzag2:center}
          \begin{minipage}[b]{0.5\textwidth}
              \centering
              \includegraphics[width=6cm,height=6cm]{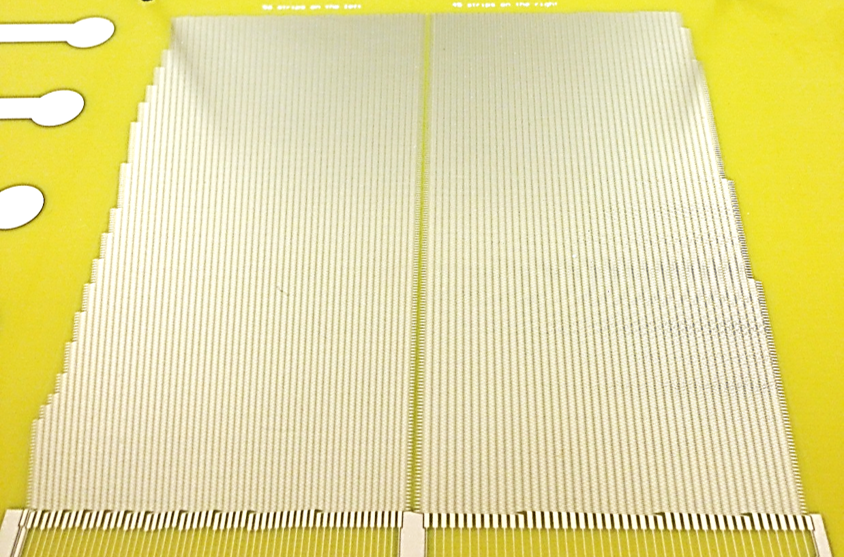}
          \end{minipage}}
    \caption{Left: The new zigzag design with improved interleaving between zigzag strips. The dashed blue lines represent the center lines of each zigzag strip. The tips of adjacent strip reach all the way to the center lines. The period of the zigzag structure in the R direction is 0.5 mm and the fraction parameter $f$ is 0.4. Right: Photo of a 10 cm $\times$ 10 cm zigzag readout area on a test board with two sections: 56 radial strips on the left with an angle pitch of 4.14 mrad and radii from 206~mm to 306~mm vs.\ 45 radial strips on the right with an angle pitch of 1.37 mrad and radii from 761~mm to 861~mm. The curved bands visible across the strips are artefacts due to digital photography.}
    \label{zigzag2}
  \end{center}
\end{figure}

Instead, we produce and test new zigzag strip boards with the ``100\% interleaving'' design~(Fig.~\ref{zigzag2}~left). For versatility, each board is designed with two different radial strip geometries. On the left side of the board, the radial strips are arranged with an angle pitch of 4.14 mrad and starting radius of 206 mm, while on the right side of the board radial strips are arranged with an angle pitch of 1.37 mrad and starting radius of 761~mm (Fig.~\ref{zigzag2}~right). We choose these two particular parameter sets because they correspond roughly to the inner and outer radial sections of a large-area trapezoidal GEM detector design for an EIC forward tracker prototype. They also have similar linear strip pitches around 1 mm. The two other design parameters, period of the zigzag structure in the R direction and fraction parameter $f$, are kept the same at 0.5 mm and 0.4, respectively. The strips are again about 10 cm long except for those near the edges of the active area. The capacitance between two adjacent zigzag strips is measured to be $(22\pm2)$~pF and the capacitance between a strip and the readout board ground is measured to be $(28\pm2)$~pF. The errors here reflect strip-to-strip variations in the measurement. 

\begin{figure}[h]
  \centering
  \includegraphics[width=16cm]{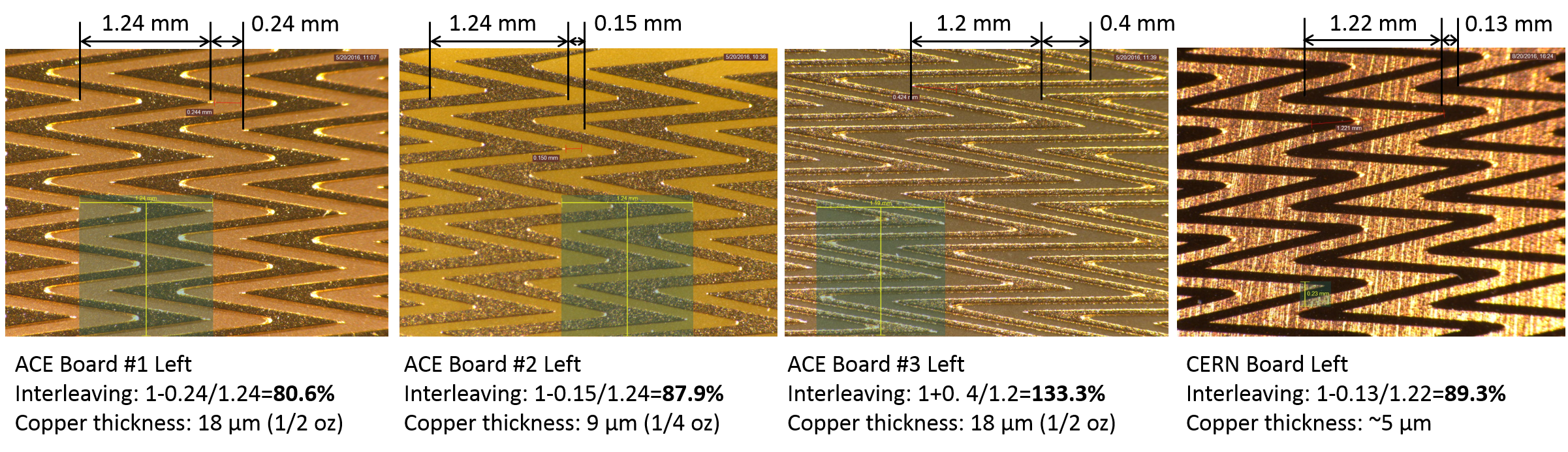}
  \vspace{-0.5cm}
  \caption{Microscopy photos of the zigzag readout structures with angle pitch 4.14 mrad produced at a PCB factory (ACE) and at CERN. Black vertical lines indicate strip pitch and distance between tips that are used to determine the amount of strip interleaving.}
  \label{ACEboardLeft}
\end{figure}

We find that the ``100\% interleaving'' design pushes the limits of industrial PCB production capabilities since the strong interleaving requires the spaces between adjacent strips to be less than 3 mils (76 $\mu$m). Three versions of the board are produced by a US PCB company (Accurate Circuit Engineering, ACE) and feature actual interleaves of 81\%, 88\%, and 133\%, respectively (Fig.~\ref{ACEboardLeft}). CERN also produced a board with the same design, but with zigzag strips  implemented on a kapton foil instead of on a PCB. The foil is then glued onto a honeycomb board for mechanical support. The main reason for producing this board is that currently CERN is the only place that can produce 1-meter-long zigzag designs on a kapton foil as required for our future R\&D demands. Consequently, this small board serves as a pilot to see whether CERN can produce the zigzag strip structures on foil with sufficient accuracy.

\begin{figure}[h]
  \centering
  \includegraphics[width=8cm]{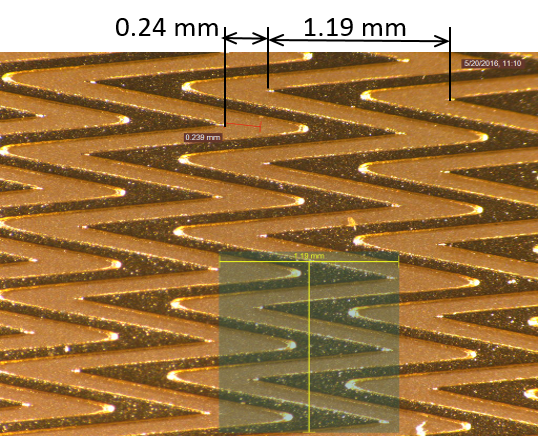}
  \caption{Microscopy photo of the zigzag readout structure in ACE\#1 board in the section with angle pitch 1.37 mrad. Here the strip interleaving is about 80\%.}
  \label{ACEboardRight}
\end{figure}

Fig.~\ref{ACEboardLeft} shows microscopy photos of the left sections of ACE and CERN boards with strip angle pitch 4.14 mrad. The first board (``ACE\#1'') is produced using a standard PCB etching method, where copper thickness is about 18 $\mu$m (1/2 oz.\ PCB standard). In this board the tips are overetched, so they are not fully reaching the adjacent strip centers. The estimated interleaving between zigzag strips is 81\%. Fig.~\ref{ACEboardRight} shows a microscopy photo of the right section of the same board with strip angle pitch of 1.37 mrad; the interleaving here is about 80\%. The second board (``ACE\#2'') is produced by using half the copper thickness (9 $\mu$m, 1/4 oz.\ PCB std.) of ACE board \#1. The resulting interleaving of 88\% comes closest to the actual design. With the third board (``ACE\#3''), we try to actively compensate in the design for the overetching problem. However, it turns out that in the actual board the strips become too thin (Fig.~\ref{ACEboardLeft}) and the interleaving comes out significantly larger than 100\%. Finally, the CERN foil board (``CERNZZ'') achieves an interleaving of 89\%, which is the closest to 100\% interleaving that has been achieved with a chemical etching technique so far. For convenience, we refer to the various boards described above for short as ZZ48, ZZ30, ACE \#1, ACE \#2, ACE \#3, and CERNZZ from here on. 

\begin{figure}[h]
  \centering
  \includegraphics[width=12cm]{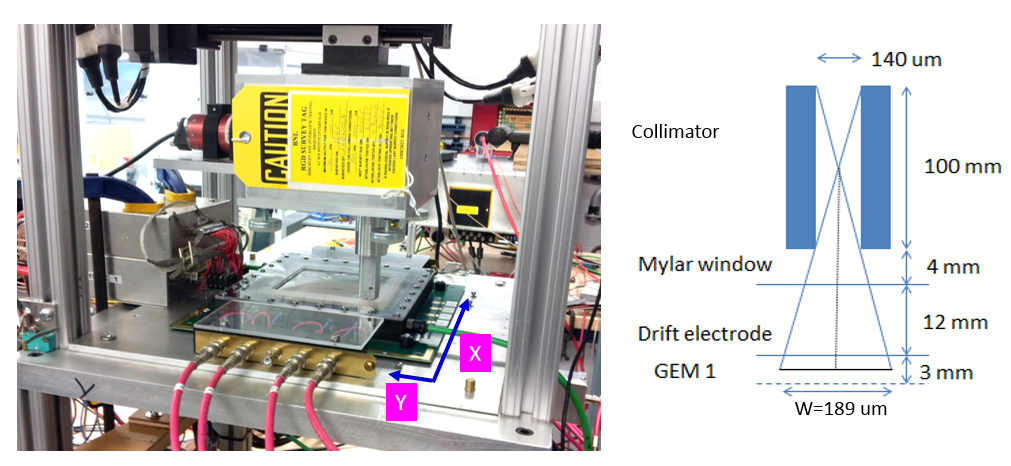}
  \vspace{-0.5cm}
  \caption{Left: Photo of the experimental setup with X-ray gun and detector mounted on a 2D stage. Right: A sketch of the X ray collimator. The collimator is 100 mm in depth with a slit 8 mm long and 140 $\mu$m wide; it is placed about 16 mm above the drift electrode so that the geometrical width ($W$) of the area that the X-rays impinge on is about 189 $\mu$m at the center of the drift region.}
  \label{setupfig}
\end{figure}

\section{Experimental configuration and procedure}\label{setup}
We test all boards with the same 10 cm $\times$ 10 cm triple-GEM detector and a highly collimated X-ray gun on a 2D motorized stage. The detector is flushed with Ar/CO$_2$ (70:30). Fig.~\ref{setupfig} shows the setup and a sketch of the X-ray collimator. In order to be consistent with our previous study~\cite{FITZZ2}, the gas gaps in the GEM detector are set to 3/1/2/1 mm for drift gap, transfer gaps 1 and 2, and induction gap, respectively. We apply high voltage settings as if there was an HV divider with a resistor chain of 1/0.5/0.5/0.45/1/0.45/0.5~M$\Omega$ so that a single voltage $V_\mathrm{drift}$ applied to the drift electrode will determine all the electric fields in the detector although individual HV channels are actually used to power the electrodes. The X-ray collimator is 100 mm thick with a slit that is 8 mm long and 140 $\mu$m wide. It is placed about 16 mm above the drift electrode and mounted on a motorized stage whose travel step length is set to 100 $\mu$m for our studies. The geometrical width of the area that the X-rays impinge on is about 189 $\mu$m at the center of the drift region (Fig.~\ref{setupfig}~right); the rate of X-rays on the detector is only about 7 Hz (not including background) due to this strong collimation. 

\begin{figure}[h]
  \centering
  \includegraphics[width=12cm]{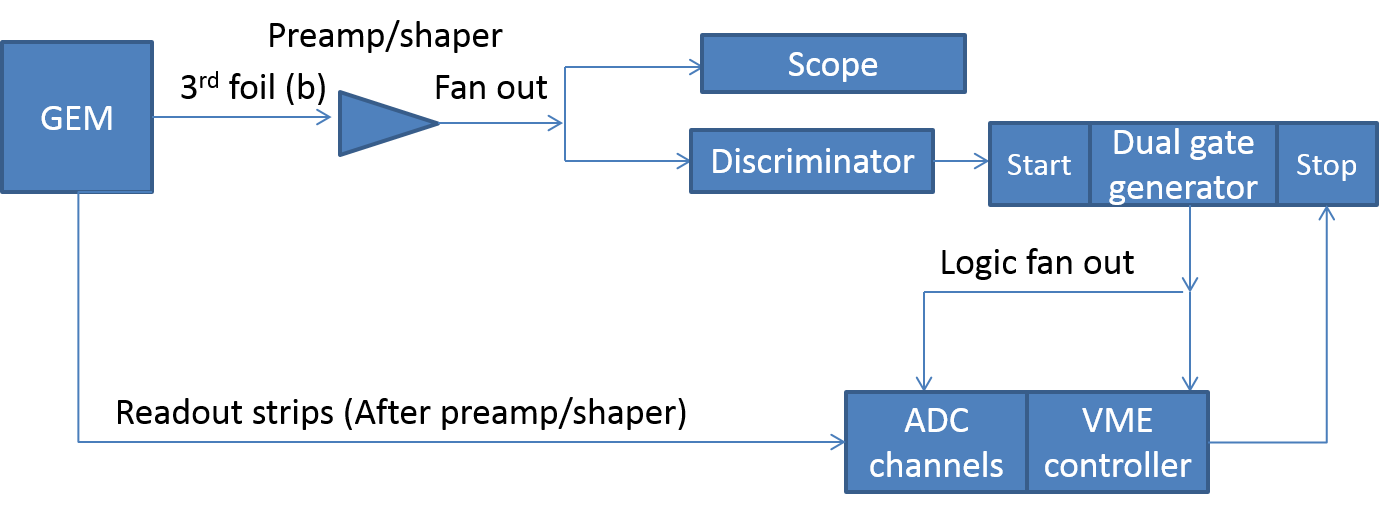}
  \vspace{-0.5cm}
  \caption{A sketch of the data aquisition electronics and logic.}
  \label{daqSketch}
\end{figure}

The data acquisition software RCDAQ was previously developed at BNL~\cite{rcdaqref} and a sketch of the electronics logic is shown in Fig.~\ref{daqSketch}. Twenty-four pairs of charge-sensitive preamplifiers and shapers are connected to 24 zigzag strip channels and are read out by three 8-channel VME Flash ADCs (Struck SIS 3300/3301) with a VME controller (CAEN 1718). The DAQ trigger is formed from the signal induced on the bottom of the third GEM. An accepted trigger stops the sampling of the FADCs, which are then read out. A signal typically occupies about 200~ns. To efficiently catch signals within the sampling time window, we adjust a window of 5 $\mu$s with a sampling rate of 100 MHz, i.e.\ we take 500 samples per window.

\begin{figure}[h]
  \centering
  \includegraphics[width=10cm]{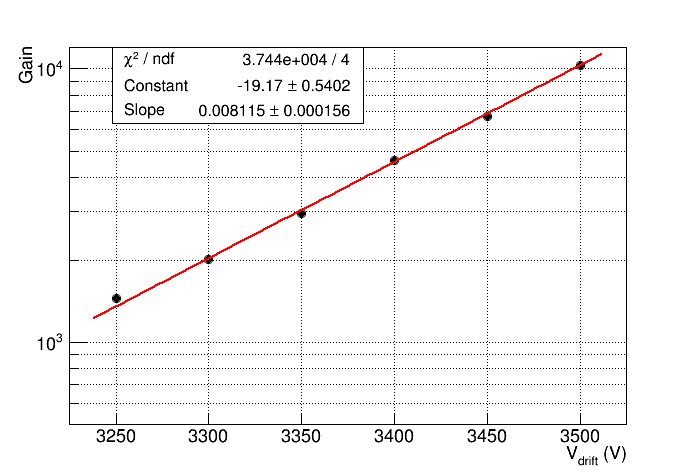}
  \vspace{-0.5cm}
  \caption{Gas gain in Ar/CO$_2$ (70:30) measured with the ACE\#1 board with 1.37 mrad strip pitch angle and an $^{55}$Fe source as a function of applied $V_\mathrm{drift}$.}
  \label{Fe55Gain}
\end{figure}

Each board is scanned in X direction across the strips as defined in Fig.~\ref{setupfig}~(left) with a step size of 100~$\mu$m over a few millimeter distance. The movement of the collimated X-Ray source closely follows the azimuthal direction. Since the zigzag strips are very close to parallel and the X-ray collimator slit has an 8~mm length, no effort is made to align the slit perfectly radially. At each point, we collect data for 8000-10000 triggered signals. The X-ray collimator is positioned so that the 140 $\mu$m slit width is oriented along the desired scan direction. The GEM detector is operated at a moderate gas gain of a few thousand; the applied $V_\mathrm{drift}$ ranges from 3250~V to 3500~V during the tests. As a reference, Fig.~\ref{Fe55Gain} shows a gain curve measured with an $^{55}$Fe source for board ACE \#1 with strip pitch angle 1.37 mrad.

\section{Experimental results}\label{results}

\subsection{Response linearity}
\begin{figure}[h]
  \begin{center}
     \subfigure{
          \label{ZZ48ZZ30-X-regions:left}
          \begin{minipage}[b]{0.5\textwidth}
              \centering
              \includegraphics[width=8cm,height=5cm]{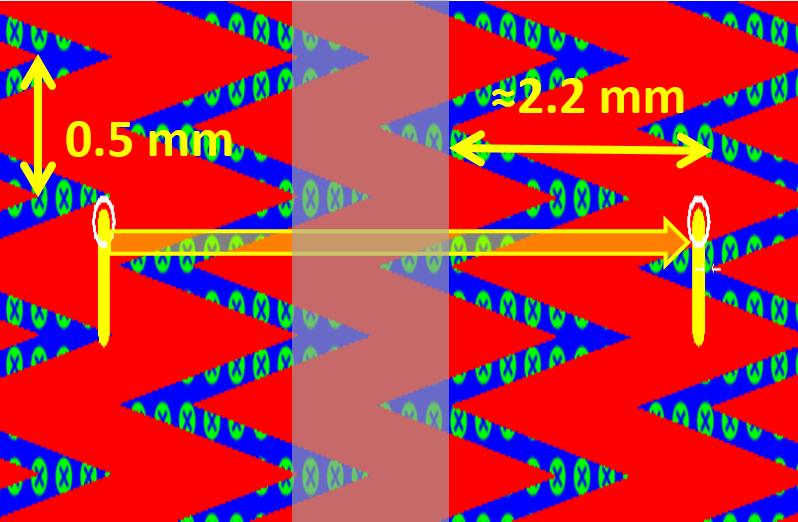}
          \end{minipage}}
    \subfigure{
          \label{ZZ48ZZ30-X-regions:right}
          \begin{minipage}[b]{0.5\textwidth}
              \centering
              \includegraphics[width=8cm,height=5cm]{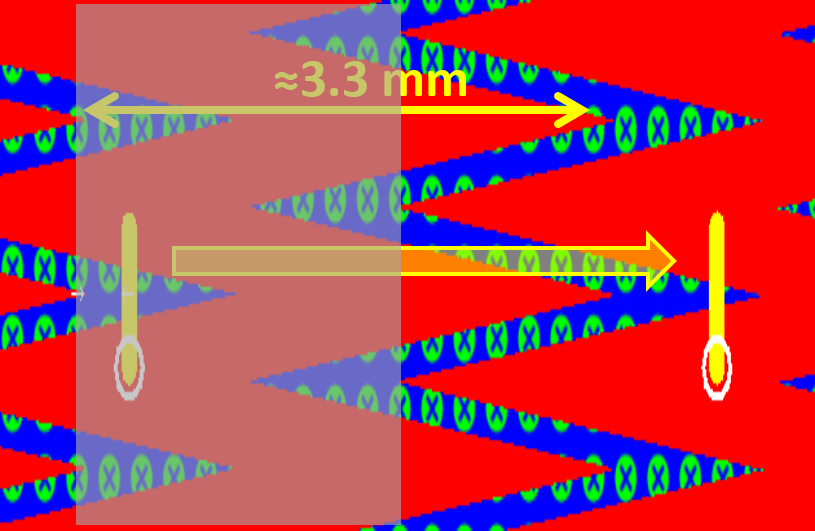}
          \end{minipage}}\newline%
    \vspace{-1cm}
    \caption{Best estimate of the scanned regions on the ZZ48 (left) and ZZ30 (right) boards relative to their design geometry. The shaded areas on both plots indicate regions that are geometrically covered by only one zigzag strip.}
    \label{ZZ48ZZ30-X-regions}
  \end{center}
\end{figure}

The ZZ48 board is scanned across a range corresponding to two strip pitches while the ZZ30 board, which has wider strips, is scanned over one strip pitch (Fig.~\ref{ZZ48ZZ30-X-regions}). For all data, we reconstruct the X-ray hit position in the X-direction in the usual fashion using the charge sharing among strips in a strip cluster. As the hit position, we take the charge-weighted centroid of the strip positions $\mathrm{s_c} = \Sigma_{i=1}^{n}\mathrm{q_i\cdot s_i} / \Sigma_{i=1}^{n} \mathrm{q_i}$, where $\mathrm{s_i}$ and $\mathrm{q_i}$ are the position at the strip center and the induced charge (in ADC counts), respectively, for the i$^\mathrm{th}$ strip in the cluster. As radial strips are intended to measure azimuthal positions, we use azimuthal strip positions in the centroid calculation by default.

\begin{figure}[h]
\vspace{-1.0cm}
  \begin{center}
     \subfigure{
          \label{ZZ48ZZ30-X-centroid:left}
          \begin{minipage}[b]{0.5\textwidth}
              \centering
              \includegraphics[width=8cm,height=5cm]{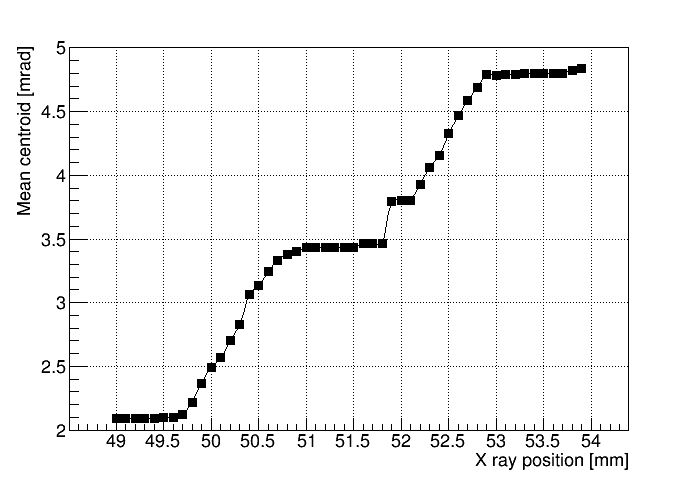}
          \end{minipage}}
    \subfigure{
          \label{ZZ48ZZ30-X-centroid:right}
          \begin{minipage}[b]{0.5\textwidth}
              \centering
              \includegraphics[width=8cm,height=5cm]{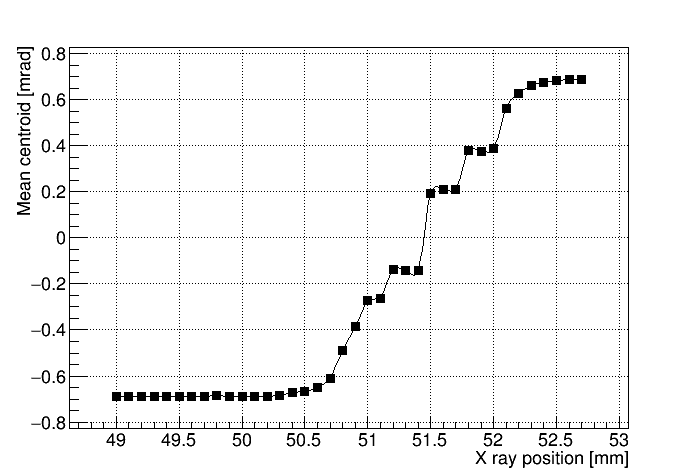}
          \end{minipage}}\newline%
    \vspace{-1cm}
    \caption{Mean strip cluster centroid vs.\ X-ray position for the ZZ48 (left) and ZZ30 (right) boards measured by scans in the direction across strips.}
    \label{ZZ48ZZ30-X-centroid}
  \end{center}
\end{figure}

\begin{figure}[h]
\vspace{-0.8cm}
  \begin{center}
     \subfigure{
          \label{ZZ48ZZ30-X-stripmulti:left}
          \begin{minipage}[b]{0.5\textwidth}
              \centering
              \includegraphics[width=8cm,height=5cm]{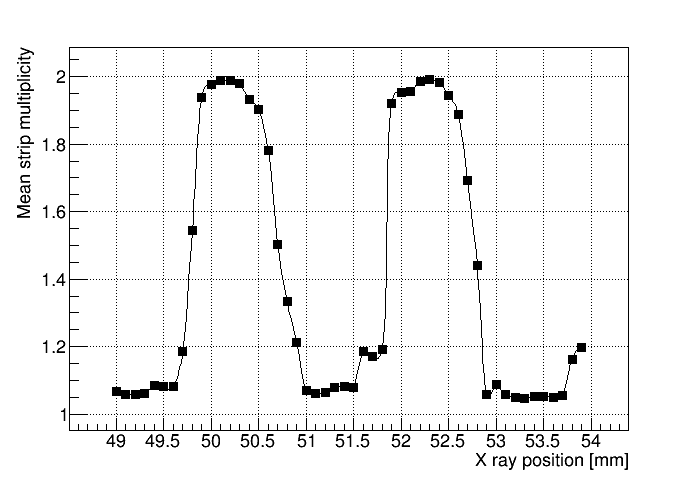}
          \end{minipage}}
    \subfigure{
          \label{ZZ48ZZ30-X-stripmulti:right}
          \begin{minipage}[b]{0.5\textwidth}
              \centering
              \includegraphics[width=8cm,height=5cm]{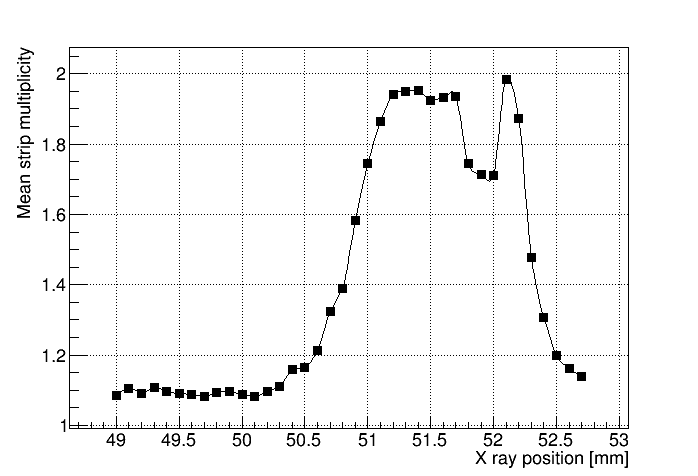}
          \end{minipage}}\newline%
    \vspace{-1cm}
    \caption{Mean strip multiplicity of strip clusters vs.\ X-ray position for the ZZ48 (left) and ZZ30 (right) boards measured by scans in the direction across strips.}
    \label{ZZ48ZZ30-X-stripmulti}
  \end{center}
\end{figure}

Fig.~\ref{ZZ48ZZ30-X-centroid} shows the mean strip-cluster centroid over all hits vs.\ actual position of X-rays incident on the boards. We observe flat regions in the curves where the detector is basically completely insensitive to the X-ray position. Checking the mean strip multiplicity of strip clusters for hits in these regions reveals that in most of these events only a single strip shows a signal (Fig.\ref{ZZ48ZZ30-X-stripmulti}). These flat regions are centered on the ``spines'' of the physical zigzag strips (see Fig.~\ref{zigzag1}). The electron avalanche is not wide enough to induce sufficient charge on adjacent strips for a good charge sharing measurement given the actual geometry of these physical zigzag strips. This contributes to an overall non-linear response in these readout structures.

The ACE and CERNZZ boards are scanned across the strips with angle pitch 4.14 mrad near a radius of 229 mm and across the strips with angle pitch 1.37 mrad near a radius of  784 mm. The scans range over 5 mm, which corresponds to 4-5 strip pitches, with a step size of 100~$\mu$m and about 8000 triggered signals per point. We focus the discussion on the results for boards ACE\#2 and CERNZZ since their zigzag structures are close to optimal and closest to the design.

Since the collimator slit is 8 mm long in the Y-direction along the strips, it exposes about 16 zigzag periods along a strip to X-rays. Consequently, our response and resolution results intrinsically incorporate any biases in the reconstruction of the X-position due to the range of X-ray hit positions along the strip in Y-direction. This bias is being minimized by design because the 500 $\mu$m zigzag period along the strip is of comparable or even smaller size (depending on the total amount of charge induced) than the diameter of the area where charge is being induced on the readout by the GEM avalanche. This means that the area of induced charge always covers at least one zigzag period along the strip and any effects due to the X-ray hit position in the radial direction along the strip average out. 

\begin{figure}[h]
  \centering
  \includegraphics[width=16cm]{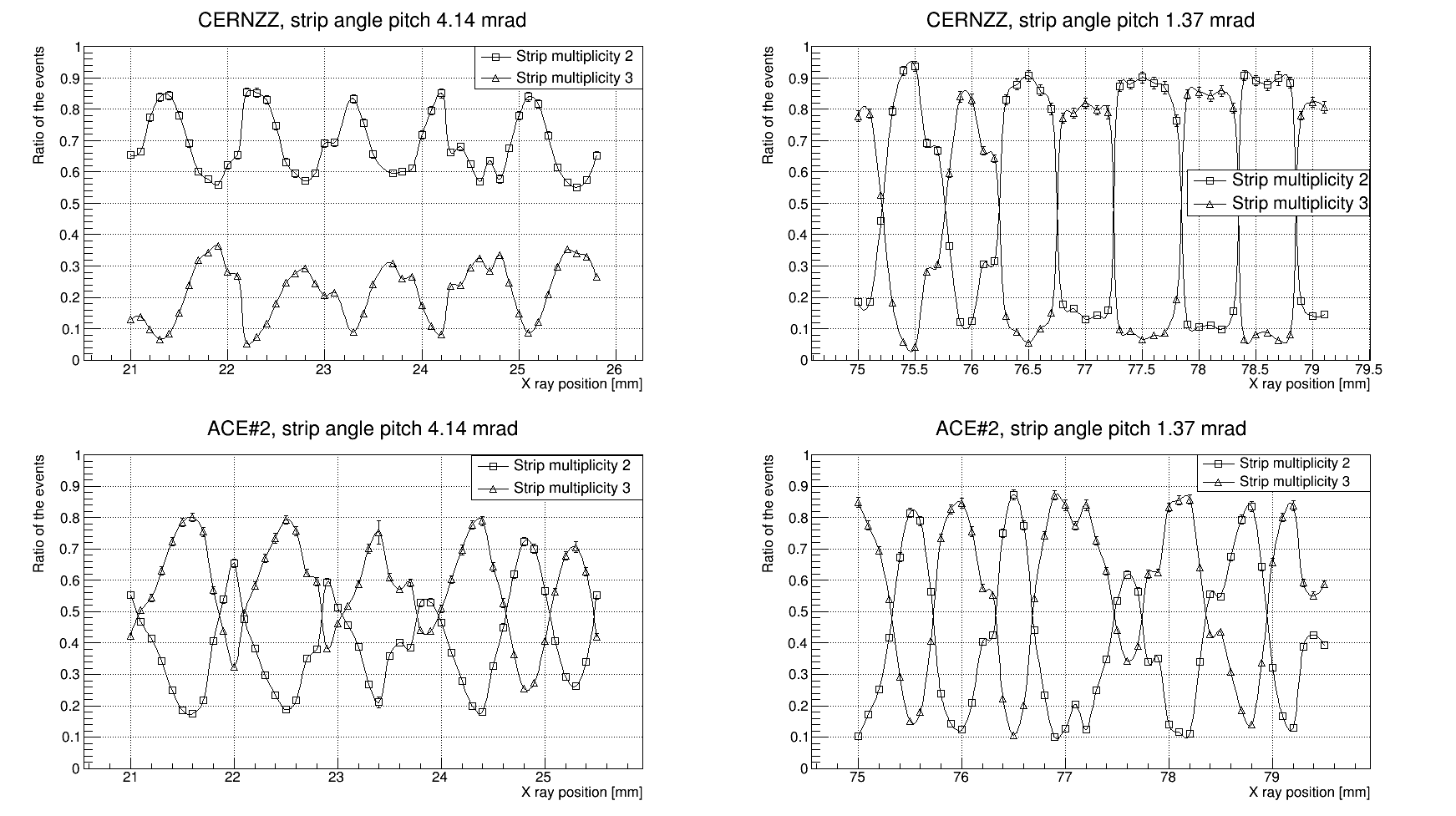}
  \vspace{-0.5cm}
  \caption{Fraction of events with strip multiplicities 2 and 3 in strip clusters observed for the CERNZZ and ACE\#2 boards for both strip angle pitches. Statistical error bars are present, but very small in the plots.}
  \label{stripMultiEvtRatio}
\end{figure}

\begin{figure}[h]
\vspace{-1cm}
  \centering
  \includegraphics[width=10cm]{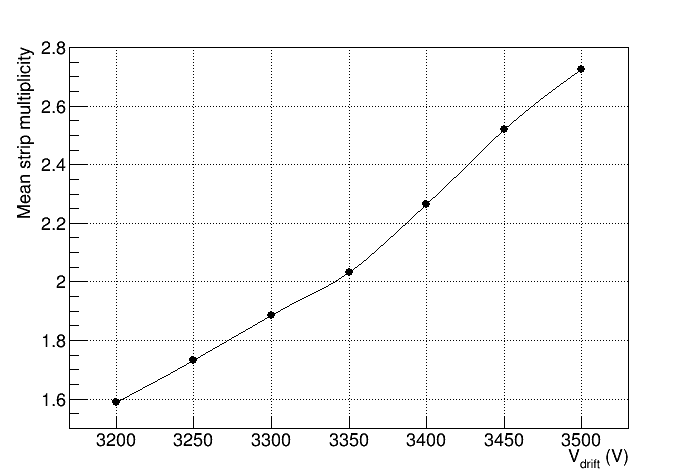}
  \vspace{-0.5cm}
  \caption{Mean strip multiplicity as a function of high voltage measured with an $^{55}$Fe source for the ACE\#1 board and 1.37 mrad strip pitch.}
  \label{StripMulti_vs_HV}
\end{figure}

As shown in Fig.~\ref{stripMultiEvtRatio}, the hits are dominated by 2-strip and 3-strip clusters, which comprise $>$ 90\% of all hits, while in the remaining hits only a single strip shows a signal. The cluster strip multiplicity is correlated with the gas gain, which is adjusted by the applied HV. The higher the gain, the smaller the population of single-strip hits. The mean cluster strip multiplicity is basically a linear function of the HV applied to the drift electrode (Fig.\ref{StripMulti_vs_HV}).

\begin{figure}[h]
  \begin{center}
     \subfigure{
          \label{newboards-xscan-stripmulti:left}
          \begin{minipage}[b]{0.5\textwidth}
              \centering
              \includegraphics[width=8cm,height=5cm]{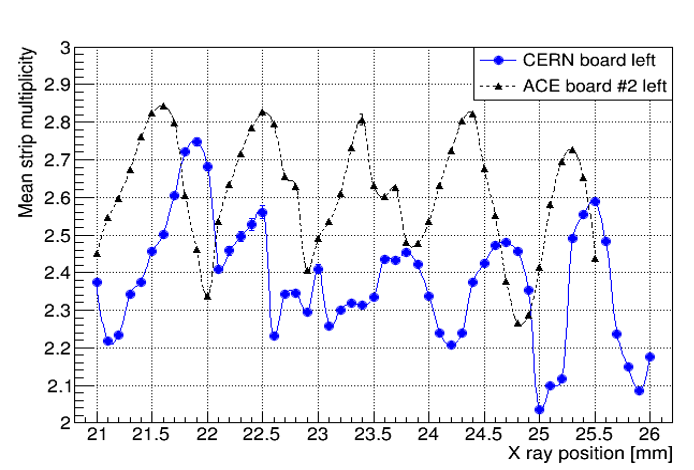}
          \end{minipage}}
    \subfigure{
          \label{newboards-xscan-stripmulti:right}
          \begin{minipage}[b]{0.5\textwidth}
              \centering
              \includegraphics[width=8cm,height=5cm]{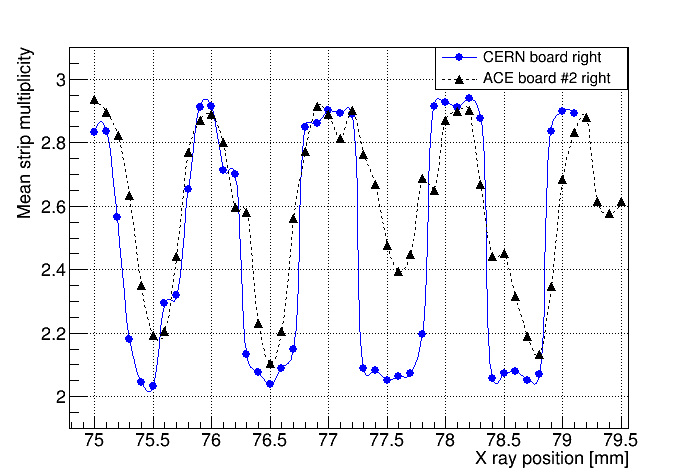}
          \end{minipage}}\newline%
    \vspace{-1cm}
    \caption{Mean strip multiplicities of strip clusters at each X-ray position for the ACE\#2 and CERNZZ boards for strips with 4.14 mrad pitch near a radius of 229 mm (left) and for strips with 1.37 mrad pitch near a radius of 784 mm (right).}
    \label{newboards-xscan-stripmulti}
  \end{center}
\end{figure}

The mean strip multiplicities as a function of X-ray position are compared for the two boards in Fig.~\ref{newboards-xscan-stripmulti}. We observe a comb-like pattern as the X-ray position moves across the zigzag structure as expected. When the X-rays hit between the centers of two strips, the mean multiplicity is close to two, whereas when they hit near the center of a strip, the multiplicity is closer to three which indicates that both adjacent strips share the charge with the central strip as intended. This effect is a bit more pronounced for the strips with 1.37 mrad angle pitch since their linear pitch of 1.07 mm is slightly larger than the 0.95 mm linear pitch for the strips with 4.14 mrad angle pitch.

\begin{figure}[h]
\vspace{-0.5cm}
  \begin{center}
     \subfigure[ACE\#2, strips with 4.14 mrad pitch near r = 229 mm]{
          \label{newboards-xscan-centroid:topleft}
          \begin{minipage}[b]{0.5\textwidth}
              \centering
              \includegraphics[width=6.5cm,height=5cm]{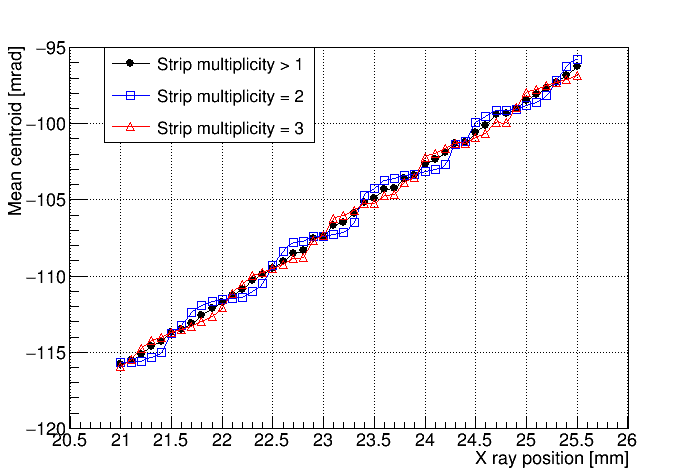}
          \end{minipage}}
     \subfigure[ACE\#2, strips with 1.37 mrad pitch near r = 784 mm]{
          \label{newboards-xscan-centroid:topright}
          \begin{minipage}[b]{0.5\textwidth}
              \centering
              \includegraphics[width=6.5cm,height=5cm]{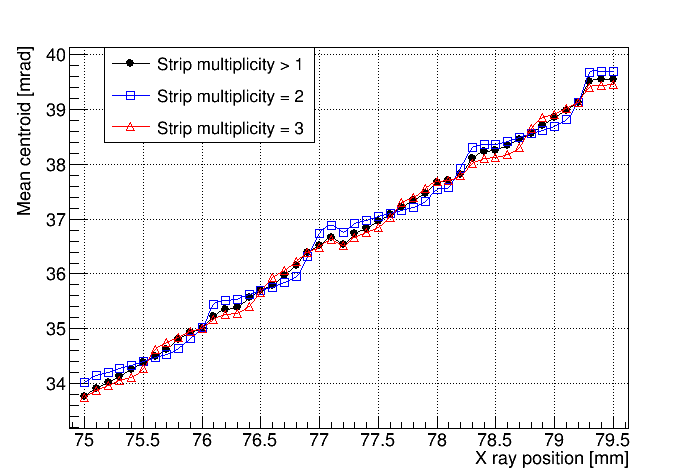}
          \end{minipage}}\newline%
    \vspace{-0.3cm}
    \subfigure[CERNZZ, strips with 4.14 mrad pitch near r = 229 mm]{
          \label{newboards-xscan-centroid:bottomleft}
          \begin{minipage}[b]{0.5\textwidth}
              \centering
              \includegraphics[width=6.5cm,height=5cm]{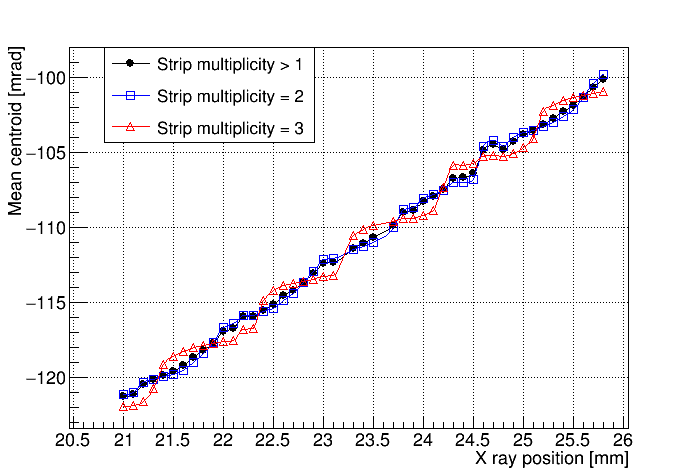}
          \end{minipage}}
    \subfigure[CERNZZ, strips with 1.37 mrad pitch near r = 784 mm]{
          \label{newboards-xscan-centroid:bottomright}
          \begin{minipage}[b]{0.5\textwidth}
              \centering
              \includegraphics[width=6.5cm,height=5cm]{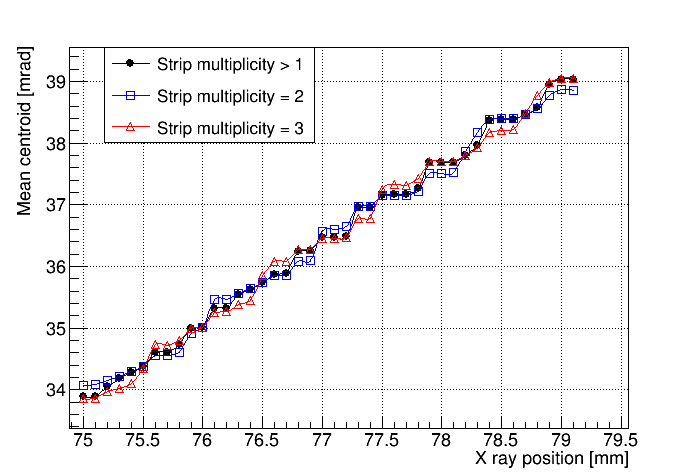}
          \end{minipage}}\newline%
    \vspace{-0.5cm}
    \caption{Mean strip-cluster centroids vs.\ X-ray positions for the ACE\#2 and CERNZZ boards. Statistical error bars are smaller than marker size in these plots.}
    \label{newboards-xscan-centroid}
  \end{center}
\end{figure}

Fig.~\ref{newboards-xscan-centroid} shows the mean strip-cluster centroid position measured at each X-ray position for boards ACE\#2 and CERNZZ. Results are compared for strip multiplicities 2, 3, and overall, but excluding single-strip clusters. The small regions in the plots with negative slopes are due to motor backlash encountered during the scans when data taking was interrupted. Comparing these responses with those shown in Fig.~\ref{ZZ48ZZ30-X-centroid} clearly demonstrates that the spatial response is much more linear for the improved ACE\#2 and CERNZZ zigzag strip boards than for the original ZZ48 and ZZ30 boards.

\subsection{Spatial resolutions}

\begin{figure}[h]
  \begin{center}
     \subfigure{
          \label{ZZ48Residual:left}
          \begin{minipage}[b]{0.33\textwidth}
              \centering
              \includegraphics[width=6cm,height=5cm]{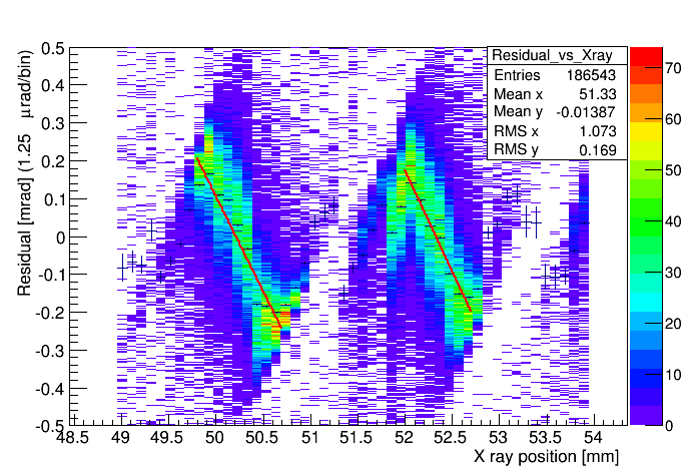}
          \end{minipage}}
    \subfigure{
          \label{ZZ48Residual:middle}
          \begin{minipage}[b]{0.33\textwidth}
              \centering
              \includegraphics[width=6cm,height=5cm]{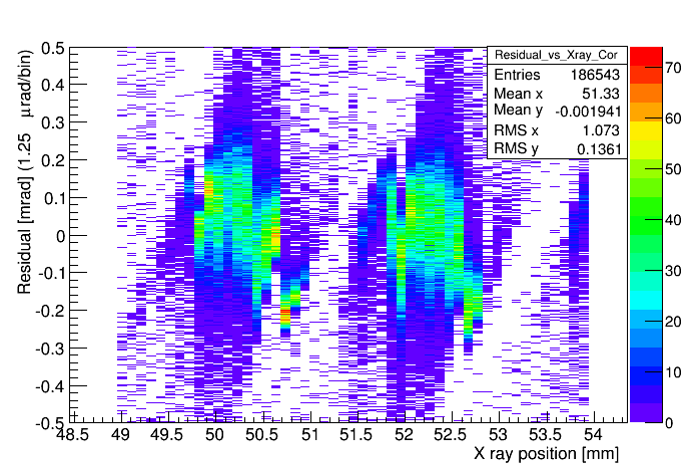}
          \end{minipage}}
    \subfigure{
          \label{ZZ48Residual:right}
          \begin{minipage}[b]{0.33\textwidth}
              \centering
              \includegraphics[width=6cm,height=5cm]{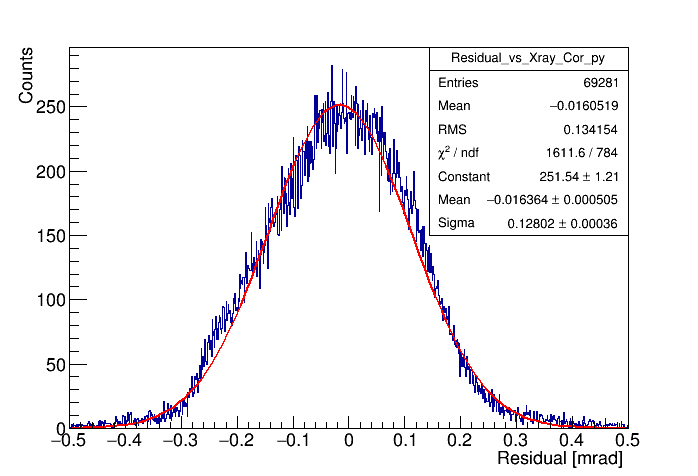}
          \end{minipage}}\newline%
    \vspace{-1cm}
    \caption{The residuals for the scans with the ZZ48 board before (left) and after (center) correction for the non-linear response. Residuals over one strip pitch from X-ray positions 51.8~mm to 52.8~mm are histogrammed in the right plot.}
    \label{ZZ48Residual}
  \end{center}
\end{figure}

A scatter plot of the residuals from the mean centroid using all cluster centroids vs.\ X-ray position for ZZ48  is shown on the left in Fig.~\ref{ZZ48Residual}. It is not a horizontal band because the response is not linear. We correct this residual distribution for the non-linear response by first fitting the mean centroid vs.\ X-ray position plot from Fig.~\ref{ZZ48ZZ30-X-centroid}~(left) with linear functions in the non-flat regions and then subtracting them from the residuals in the scatter plots to flatten them out. This pushes the band of residuals towards a more horizontal line as shown in the center of Fig.~\ref{ZZ48Residual}. Projecting this plot onto the vertical residual axis over one strip pitch produces the distribution shown on the right in Fig.~\ref{ZZ48Residual}. The width of this distribution, which we use as a measure for the overall residual width, is 128~$\mu$rad or 188~$\mu$m at the radius of 1468~mm for the ZZ48 board.

\begin{figure}[h]
  \vspace{-1cm}
  \begin{center}
    \subfigure{
          \label{xraysim:right}
          \begin{minipage}[b]{0.5\textwidth}
              \centering
              \includegraphics[width=10cm,height=8cm]{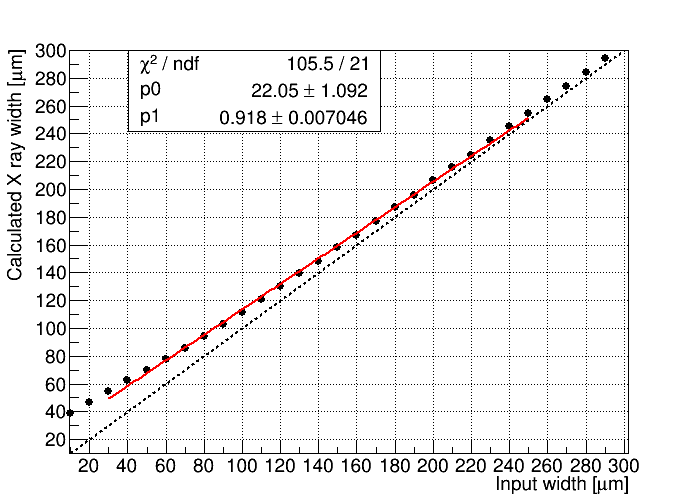}
          \end{minipage}}\newline%
    \vspace{-1cm}
    \caption{Geant4 simulation of the effect of X-ray collimation on the measurement of detector resolutions; statistical errors are smaller than marker size.}
    \label{xraysim}
  \end{center}
\end{figure}

In order to find the intrinsic spatial resolutions for the tested zigzag boards, we need to subtract the effect of the finite width of the X-ray collimator, which causes a smearing of the X-ray incidence position on the detector. This is done using a simple Geant4 simulation and the result is shown in Fig.~\ref{xraysim}. Through the simulation, the relation between an intrinsic detector resolution that is fed into the simulation and a measured width of the X-ray spot on the detector is determined. This mostly linear function is then used to obtain the intrinsic spatial resolution of the detector from the measured raw residual width observed in the data. The finite collimator width has an impact mainly below 100 $\mu$m. A few different X-ray beam shapes emerging from the source (uniform rectangular, cone, Gaussian) are tested in the simulation and their impact is found to be negligible. From the residual width and the GEANT4 curve, the intrinsic resolution of the ZZ48 board is measured to be (123.1$\pm$0.4)~$\mu$rad or (180.7$\pm$0.6)~$\mu$m after the X-ray collimator effect is subtracted. 

In our previous study~\cite{FITZZ2}, a similar zigzag readout structure was tested using hadron beams and the resolution was found to be around 180~$\mu$rad at a slightly lower voltage of V$_\mathrm{drift}=3200$~V. We estimate from the measurement of resolution as a function of HV in that previous study that a 140 $V$ increase in V$_\mathrm{drift}$ corresponds to a 27\% improvement in resolution, i.e.\ at V$_\mathrm{drift}=3340$~V the resolution would be expected to be about 130~$\mu$rad. Our  current measurement of 123~$\mu$rad with X-rays is reasonably consistent with this expectation from our previous measurement with hadrons.

\begin{figure}[h]
\vspace{-2cm}
  \begin{center}
     \subfigure{
          \label{CERNZZ-right-residual:topleft}
          \begin{minipage}[b]{0.33\textwidth}
              \centering
              \includegraphics[width=5.5cm,height=5cm]{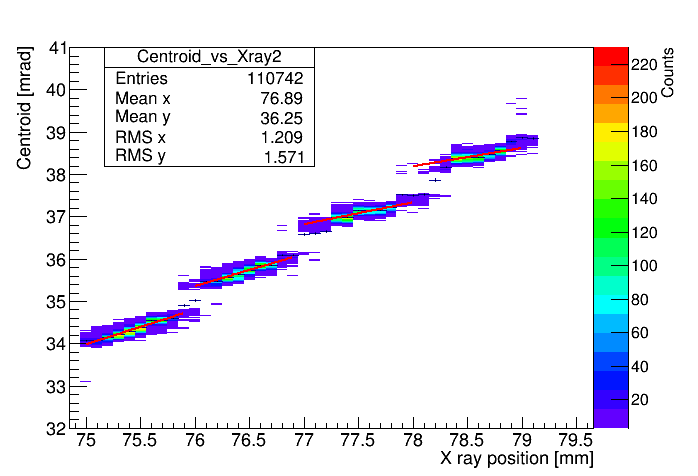}
          \end{minipage}}
     \subfigure{
          \label{CERNZZ-right-residual:topmiddle}
          \begin{minipage}[b]{0.33\textwidth}
              \centering
              \includegraphics[width=5.5cm,height=5cm]{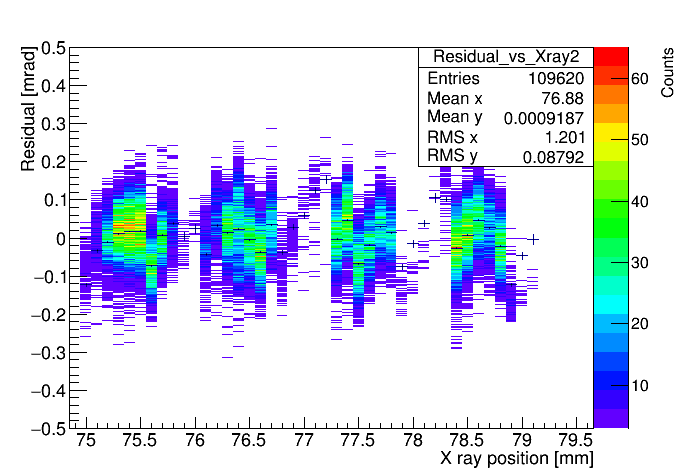}
          \end{minipage}}
     \subfigure{
          \label{CERNZZ-right-residual:topright}
          \begin{minipage}[b]{0.33\textwidth}
              \centering
              \includegraphics[width=5.5cm,height=5cm]{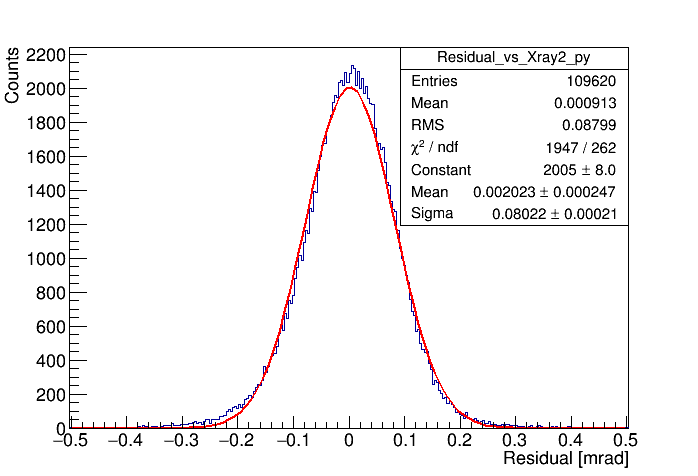}
          \end{minipage}}\newline%
    \vspace{-0.8cm}
    \subfigure{
          \label{CERNZZ-right-residual:bottomleft}
          \begin{minipage}[b]{0.33\textwidth}
              \centering
              \includegraphics[width=5.5cm,height=5cm]{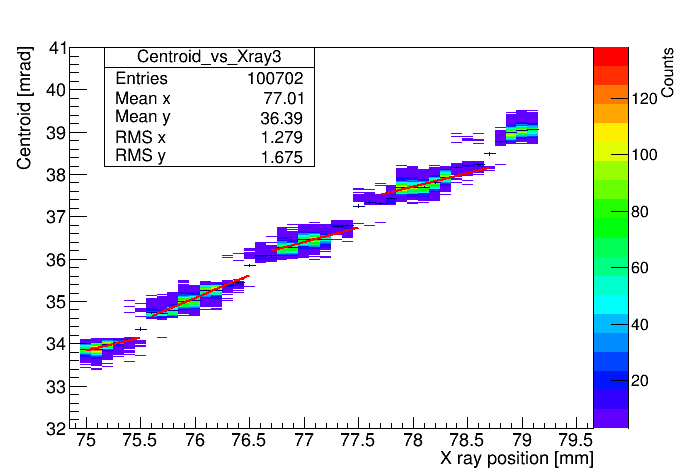}
          \end{minipage}}
     \subfigure{
          \label{CERNZZ-right-residual:bottommiddle}
          \begin{minipage}[b]{0.33\textwidth}
              \centering
              \includegraphics[width=5.5cm,height=5cm]{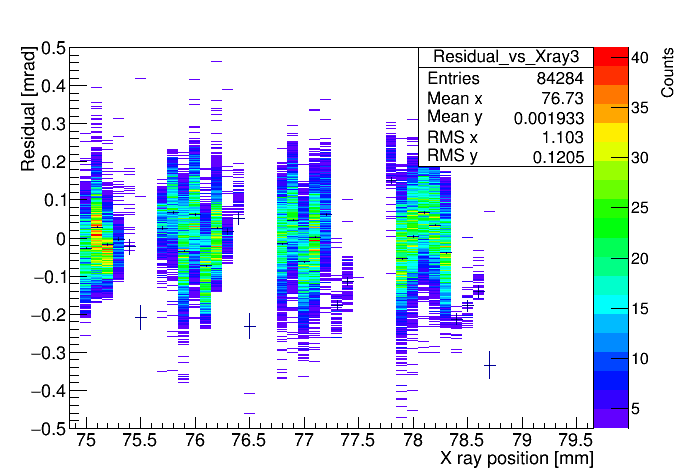}
          \end{minipage}}
      \subfigure{
          \label{CERNZZ-right-residual:bottomright}
          \begin{minipage}[b]{0.33\textwidth}
              \centering
              \includegraphics[width=5.5cm,height=5cm]{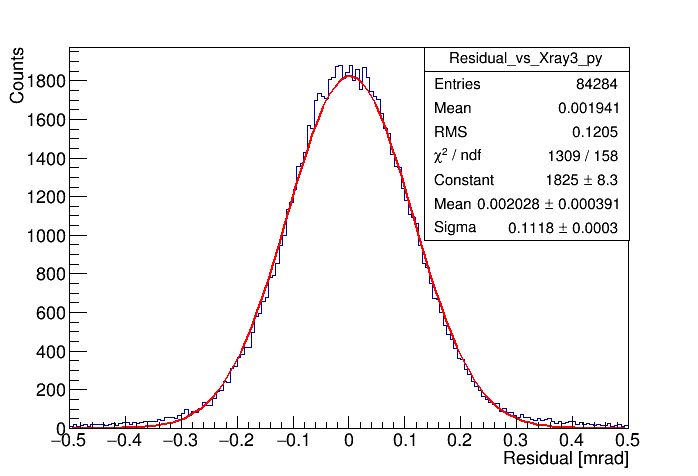}
          \end{minipage}}\newline%
    \vspace{-0.8cm}
    \caption{Residual plots for the CERNZZ board for strips with 1.37 mrad angle pitch for 2-strip clusters (top) and 3-strip clusters (bottom). Left: Cluster centroids vs.\ X-ray positions with linear fits. Center: Flattened residual scatter plots. Right: Projected residual distributions.}
    \label{CERNZZ-right-residual}
  \end{center}
\end{figure}

The same procedure is applied to the data for the other boards. Fig.~\ref{CERNZZ-right-residual} shows the residual scatter plots for 2-strip and 3-strip clusters using the strips with 1.37 mrad angle pitch on the CERNZZ board. Linear functions are fitted to the strip cluster centroid vs.\ X-ray position plots and subtracted from the residuals to flatten them out so that overall residuals can be calculated. The residual widths are found to be 80~$\mu$rad (2-strip clusters) and 112~$\mu$rad (3-strip clusters). By again subtracting the X-ray collimator effect, the corresponding measured intrinsic resolutions are 57~$\mu$rad and 92~$\mu$rad for 2-strip and 3-strip clusters, respectively. This corresponds to linear instrinsic resolutions of 45~$\mu$m and 72~$\mu$m, respectively, at a radius of 784~mm. The overall intrinsic resolution is 71~$\mu$rad or 56~$\mu$m at a radius of 784~mm if we combine 2-strip and 3-strip clusters in the analysis. Statistical errors in these analyses are less than 0.3\%.

\begin{table}[h]
\centering
\caption{Spatial resolutions for all boards with improved zigzag strip designs measured in X-ray scans. Effects due to finite X-ray collimator width are subtracted. Statistical errors are less than 0.3\%. The voltage applied to the drift electrode and approximate gas gain are listed for reference. The gain variation is estimated to be $\approx$ 30\% based on signal amplitudes observed with different boards.}\vspace{2mm}
\label{restable}
\begin{tabular}{|p{1.3cm}|p{1.1cm}|p{1.1cm}||p{1.4cm}|p{1.5cm}|p{1.6cm}|p{1.4cm}|p{1.5cm}|p{1.6cm}|}
\hline
\multicolumn{3}{|p{3.4cm}||}{\textbf{Spatial resolutions} ($\mu$rad / $\mu$m)} & 
\multicolumn{3}{p{4.5cm}|}{Strips with an angle pitch of 4.14 mrad, R $\approx$ 229 mm} & 
\multicolumn{3}{p{4.5cm}|}{Strips with an angle pitch of 1.37 mrad, R $\approx$ 784 mm} \\ \hline\hline
Board name & V$_\mathrm{drift}$ (V) & Gas gain & 2-strip clusters & 3-strip clusters & 2\hspace{0.5mm}\&\hspace{0.5mm}3-strip clusters & 2-strip clusters & 3-strip clusters & 2\hspace{0.5mm}\&\hspace{0.5mm}3-strip clusters \\ \hline
ACE\#1   & 3380 & 4000 & 266 / 61 & 371 / 85   & 328 / 75 & 56 / 44 & 69 / 54    & 60 / 47 \\ \hline
ACE\#2   & 3340 & 3000 & 288 / 66 & 480 / 110  & 384 / 88 & 57 / 45 & 97 / 76    & 84 / 66 \\ \hline
ACE\#3   & 3250 & 1500 & -        & 572 / 131  & -        & -       & 140 / 110  & - \\ \hline
CERNZZ   & 3340 & 3000 & 397 / 91 & 393 / 90   & 397 / 91 & -       & -          & - \\ \hline
CERNZZ   & 3380 & 4000 & -        & -          & -        & 57 / 45 & 92 / 72    & 71 / 56 \\ \hline
\end{tabular}
\end{table}

Table~\ref{restable} summarizes the final intrinsic resolution results for all the ACE boards and the CERNZZ board after X-ray width effects are subtracted. Most measured resolutions are better than 100~$\mu$m. The applied V$_\mathrm{drift}$ and corresponding gas gain are also listed in the table for reference. The uncertainty on the gain is estimated to be about 30\% based on observed variations in signal amplitudes during the different measurements. The CERNZZ and ACE\#2 boards show almost the same resolutions for the strips with 1.37 mrad angle pitch; the resolution difference for strips with 4.14 mrad is likely due to gain variations while scanning. The consistently better resolutions for board ACE\#1 relative to board ACE\#2 are due to the higher gain applied to ACE\#1 during the test. 

\clearpage
\section{Summary and conclusions} \label{basic_results}
The physical readout boards with improved radial zigzag strip design, which we have tested with highly collimated X-rays, feature close-to-optimal (89\%) interleaving of adjacent strips. This enables robust charge sharing among adjacent strips, which in turn results in a marked improvement in spatial response linearity and spatial resolution over previous zigzag strip designs that we have studied. The spatial detector responses are almost fully linear, so there is no need for any corrections when reconstructing hit positions from straight-forward strip cluster centroids. With linear strip pitches around 1 mm, the boards achieve linear spatial resolutions of 50-90 $\mu$m at medium gas gains around 3000. The successful production of these improved radial zigzag strips on a foil rather than on a PCB and the observation that their performance is comparable to PCBs are promising for using zigzag-strip readouts on large foils to reduce overall material of a GEM detector. Consequently, this approach lends itself to GEM detector applications where material budgets can be critical, such as TPCs and forward tracking detectors at the EIC, while reducing the overall number of required readout channels and maintaining excellent spatial resolution.

\section*{Acknowledgments}
This work is supported by Brookhaven National Laboratory under the EIC eRD-6 consortium. We also acknowledge Alexander Kiselev (BNL) for his help with finalizing the design idea for the optimized zigzag structure.

\end{document}